\documentclass[a4paper,11pt]{article}
\usepackage{bm,color}
\usepackage{pos}
\usepackage{tikz}
\usepackage{dirtree}
\hypersetup{colorlinks=true, urlcolor=cyan, linkcolor=cyan}

\newcommand{\smo}{\textsf{SModelS}}
\newcommand{\smotools}{\textsf{SModelS Tools}}
\newcommand{\json}{\textsf{JSON}}
\newcommand{\pyhf}{\textsf{pyhf}}
\newcommand{\python}{\textsf{python}}

\newcommand{\met}{E_T^{\rm miss}}
\newcommand{\BR}{{\rm BR}}

\title{New developments in \smo}

\author*[a,b]{Ga\"el Alguero}
\author[c]{Jan Heisig}
\author[d]{Charanjit K. Khosa}
\author*[a]{Sabine Kraml}
\author[e]{Suchita Kulkarni}
\author[f]{Andre Lessa}
\author[g]{Philipp Neuhuber}
\author[a,d]{Humberto Reyes-Gonz\'{a}lez}
\author[g,h]{Wolfgang Waltenberger}
\author[i]{Alicia Wongel}

\affiliation[a]{Laboratoire de Physique Subatomique et de Cosmologie, Université Grenoble-Alpes, CNRS/IN2P3,\\53 Avenue des Martyrs, F-38026 Grenoble, France}
\affiliation[b]{LAPTh, Universit\'e Savoie Mont Blanc, CNRS, B.P. 110, F-74941 Annecy Cedex, France}
\affiliation[c]{Centre for Cosmology, Particle Physics and Phenomenology (CP3), Universit\'e catholique de Louvain,\\ Chemin du Cyclotron 2, B-1348 Louvain-la-Neuve, Belgium}
\affiliation[d]{Department of Physics, University of Genova, Via Dodecaneso 33, 16146 Genova, Italy}
\affiliation[e]{Institute of Physics, NAWI Graz, University of Graz, Universit\"atsplatz 5, A-8010 Graz, Austria}
\affiliation[f]{Centro de Ciências Naturais e Humanas, Universidade Federal do ABC, Santo Andr\'e,\\ 09210-580 SP, Brazil}
\affiliation[g]{Institut f\"ur Hochenergiephysik, \"Osterreichische Akademie der Wissenschaften, Nikolsdorfer Gasse 18, A-1050 Wien, Austria}
\affiliation[h]{University of Vienna, Faculty of Physics, Boltzmanngasse 5, A-1090 Wien, Austria}
\affiliation[i]{DESY, Notkestraße 85, 22607 Hamburg, Germany}

\emailAdd{gael.alguero@lpsc.in2p3.fr}
\emailAdd{jan.heisig@uclouvain.be}
\emailAdd{charanjit.kaur@edu.unige.it}
\emailAdd{sabine.kraml@lpsc.in2p3.fr}
\emailAdd{suchita.kulkarni@uni-graz.at}
\emailAdd{andre.lessa@ufabc.edu.br}
\emailAdd{ph.neuhuber@gmail.com}
\emailAdd{humbertoalonso.reyesgonzlez@edu.unige.it}
\emailAdd{wolfgang.waltenberger@oeaw.ac.at}
\emailAdd{alicia.wongel@desy.de}

\abstract{\smo\ is an automatized tool enabling the fast interpretation of simplified model results from the LHC within any model of new physics respecting a $\mathbb{Z}_2$ symmetry. In this contribution, we report on two important updates of \smo\ during 2020: the extension of the \smo' database with 13 ATLAS and 10 CMS analyses, including 5 ATLAS and 1 CMS analyses at full Run~2 luminosity, and the ability to use full likelihoods now provided by ATLAS in the form of pyhf JSON files. 
Moreover, we briefly explain how to use \smo\ and give an overview of ongoing developments.}

\FullConference{%
  Tools for High Energy Physics and Cosmology - TOOLS2020\\
  2-6 November, 2020\\
  Institut de Physique des 2 Infinis (IP2I), Lyon, France}

\begin{document}
\maketitle

\section{Introducing SModelS}

Most experimental searches for new physics at the LHC interpret their data in the context of simplified models. Such simplified models reduce a full theory to only a few new  particles with respect to the Standard Model (SM) and assume a simple decay pattern. 
For example, the supersymmetry (SUSY) motivated search in the $2\tau+\met$ final state may typically be interpreted in a simplified model that comprises a stau, $\tilde\tau_1$, and a neutralino, $\tilde{\chi}_1^0$, as the only new particles and that considers just stau-pair production and a 100\% branching ratio for $\tilde\tau_1 \rightarrow \tau \tilde{\chi}_1^0$.
There is a plethora of ATLAS and CMS analyses from Run~1 and Run~2, looking for new physics in different final states and/or exploiting different analysis techniques, that report their results in the context of such simplified models. 

To work out how all these experimental results constrain full models, which may have a large particle content and (much) more complicated decay patterns, is a highly non-trivial task. It may be addressed by reproducing the experimental analyses in Monte Carlo event simulations---this approach, often referred to as ``recasting'', is however extremely CPU time consuming.  Moreover, only a fraction of the experimental searches is implemented in public 
recasting tools. The alternative is to re-use the published simplified model results -- this is the purpose of  \smo~\cite{Kraml:2013mwa,Ambrogi:2017neo,Ambrogi:2018ujg,Dutta:2018ioj,Khosa:2020zar,Alguero:2020grj}. This approach requires to compute only cross sections times branching ratios, $\sigma\times {\rm BRs}$, for the tested signals. It is often more conservative but \emph{much} faster than full recasting.
Besides speed, the power of \smo\ comes from its large database of experimental results, which allows to check limits from nearly 100 ATLAS and CMS searches for SUSY\@. 
Note that \smo\ can conveniently be used beyond SUSY for 
any Beyond the Standard Model (BSM) scenario with a $\mathbb{Z}_2$-like symmetry, for which the signal acceptance of the SUSY searches apply~\cite{Kraml:2013mwa}. 

In the following, we first present in more detail the working principle of \smo. This is followed by the new developments published in 2020 (\smo\ versions 1.2.3 and 1.2.4), namely a major database update \cite{Khosa:2020zar} and the \pyhf\ support~\cite{Alguero:2020grj} for usage of full likelihoods provided by ATLAS. We then briefly explain how to use \smo. 
We conclude with a short summary, some remarks on the treatment of long-lived particles, and a brief discussion of ongoing developments for the upcoming \smo\ version 2.0.
An overview of other approaches and recasting tools can be found in \cite{Abdallah:2020pec} and in the review talk by B.~Fuks at this workshop.

\subsection{Working principle}

\begin{figure}[t]    \centering
    \includegraphics[height=0.35\textheight,width=0.7\textwidth]{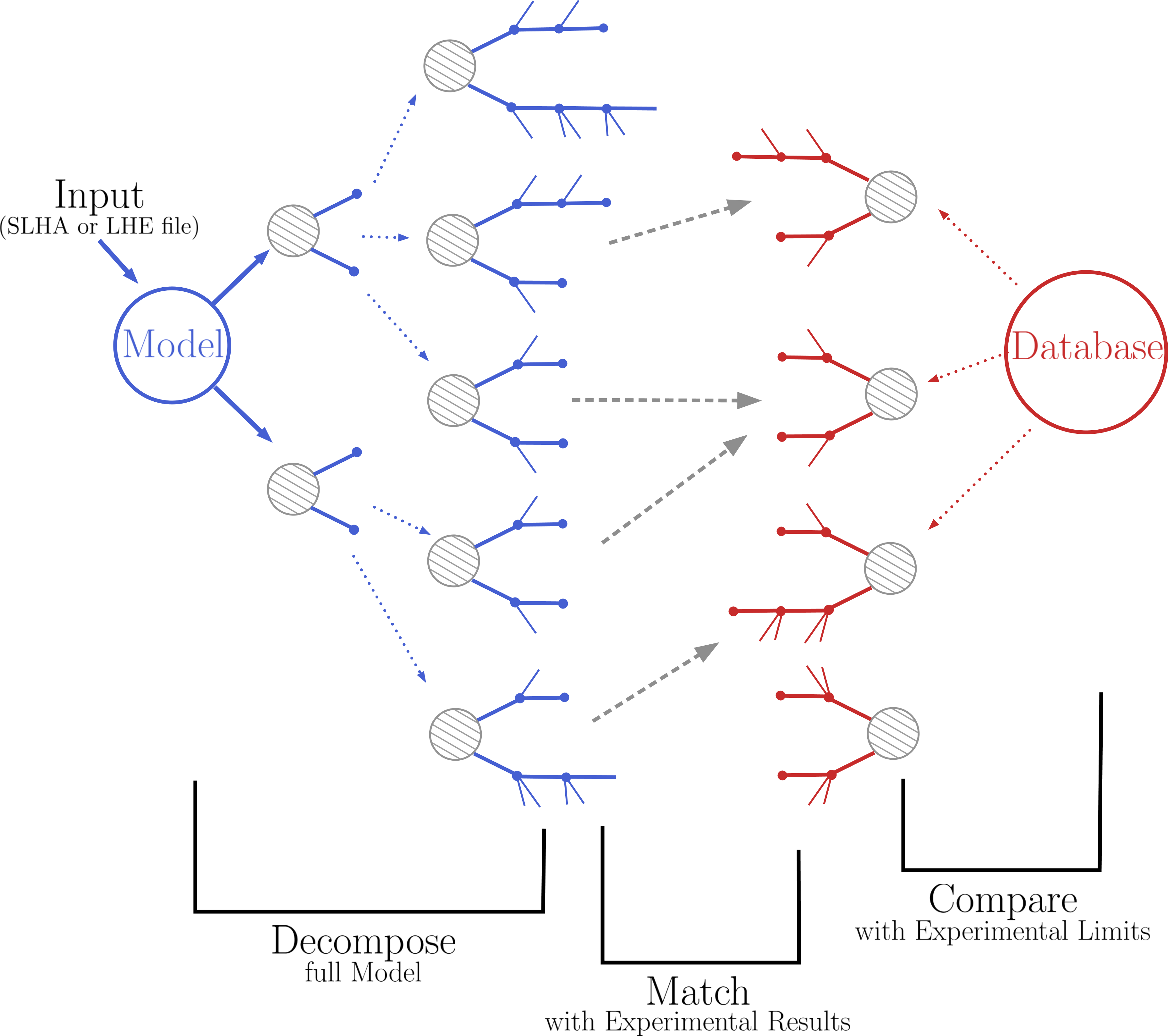}
    \caption{\smo\ working principle.}
    \label{fig:scheme}
\end{figure}

\smo\ is based on a general procedure to decompose BSM scenarios featuring a $\mathbb{Z}_2$-like symmetry into simplified-model topologies---also referred to as simplified model spectra (SMS). The decomposed topologies are then matched and constrained with experimental results in the \smo' database (from currently 48 ATLAS and 47 CMS analyses). The working principle is illustrated in Figure~\ref{fig:scheme}.

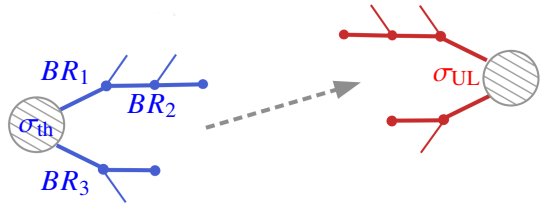
\begin{figure}[t]
    \centering
    \begin{tikzpicture}
        \node[anchor=south west,inner sep=0] (image) at (0,0) {\includegraphics[width=0.5\textwidth]{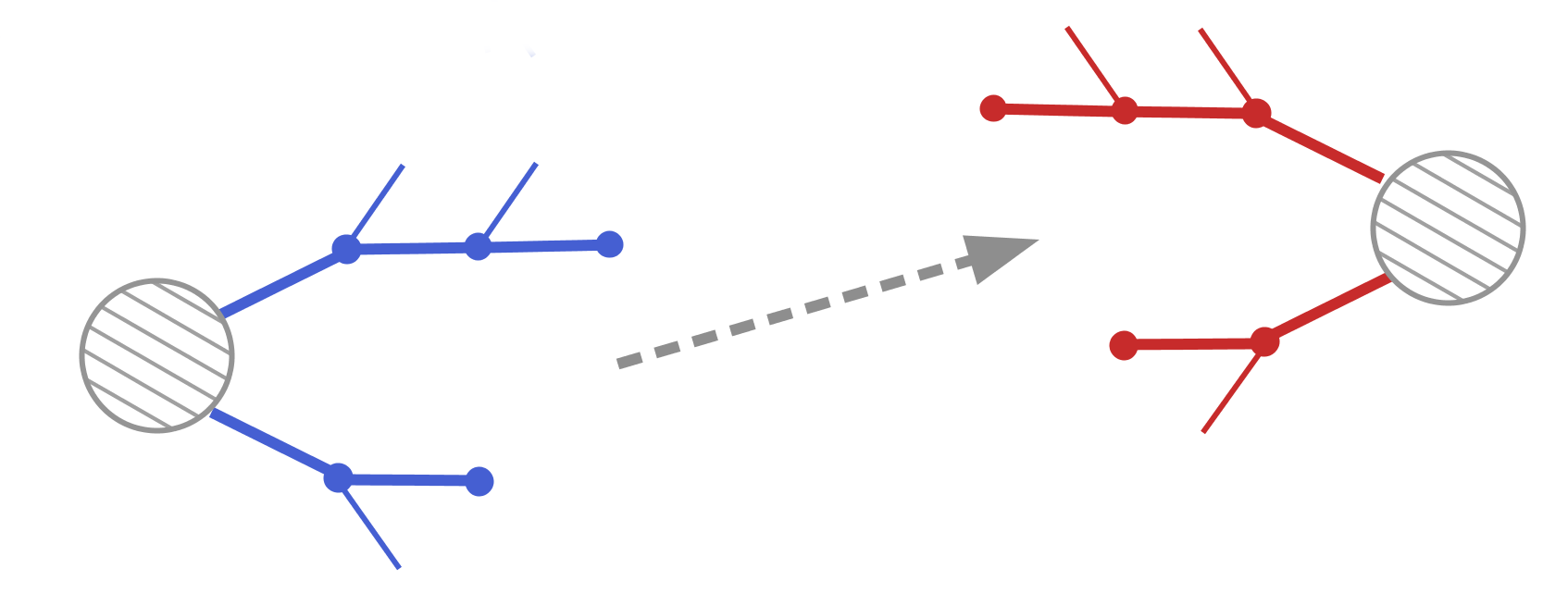}};
        \begin{scope}[x={(image.south east)},y={(image.north west)}]
            \draw[blue] (0.1, 0.4) node {$\sigma_{\rm{th}}$};
            \draw[blue] (0.15, 0.65) node {$BR_1$};
            \draw[blue] (0.15, 0.15) node {$BR_3$};
            \draw[blue] (0.3, 0.5) node {$BR_2$};
            \draw[red]  (0.83, 0.63) node {$\sigma_{\rm{UL}}$};
        \end{scope}
    \end{tikzpicture}    
    \caption{Matching procedure: $\sigma_{\rm{UL}}$ sets a limit on $\sigma_{\rm{th}} \times BR_1 \times BR_2 \times BR_3$}
    \label{fig:matching}
\end{figure}

Starting from the BSM particle masses, decay branching ratios and production cross sections provided in the input SLHA or LHE file, \smo\ determines the \emph{weights} in terms of cross sections times branching ratios, $\sigma\times\BR$, of all occurring BSM signals. These signal weights are then confronted with the experimental constraints in the database of LHC results as exemplified in Figure~\ref{fig:matching}.
This is easier and much faster than reproducing analyses with Monte Carlo event simulation, and it allows for reinterpreting searches which are not just cut and count, e.g., analyses which rely on BDT (boosted decision tree) variables.
The downside is that the applicability is limited by the simplified model results available in the database (mostly simple, symmetric signal topologies).
Moreover, whenever the tested signal splits up into many different channels, as it is often the case in complex models with many new particles, the derived limits tend to be highly conservative.  

\smo\ is thus particularly useful for evaluating constraints and generally characterizing collider signatures in 
large scans and model surveys. 
As mentioned, it can be used for any BSM scenario  (beyond the minimal supersymmetric SM, MSSM, and beyond SUSY in general) with a $\mathbb{Z}_2$-like symmetry, as long as the simplified model assumptions apply. This means, the kinematic distributions -- and thus the signal acceptance -- of the tested signal need to be approximately the same as for the (SUSY) scenario assumed for the simplified model in the experimental analysis;\footnote{In particular, the framework does not distinguish between different production modes (e.g.~$t$-channel versus $s$-channel) or different spins (see, however, section~\ref{summary} for upcoming developments).\label{fn:assump}}
see~\cite{Kraml:2013mwa} for a discussion of the simplified model assumptions and associated caveats as well as~\cite{Edelhauser:2014ena,Edelhauser:2015ksa,Kraml:2016eti} for a quantification of their impact.

\subsection{Database structure}
\label{sec:dbStruct}

The \smo\ database makes use of two types of results: {\bf upper limit (UL)} results and {\bf efficiency map (EM)} results. UL maps are 95\% confidence level (CL) upper limits on $\sigma\times\BR$ mapped on the parameter space of the given simplified model, typically the BSM masses or slices over the BSM masses. Coming back to the example of the stau-neutralino simplified model (also called TStauStau), the UL map would provide the 95\% CL limit on $\sigma(pp \rightarrow \tilde\tau_1\bar{\tilde\tau}_1)\times [\BR(\tilde\tau_1 \rightarrow \tau \tilde{\chi}_1^0)]^2$ in bins of  ($m_{\tilde{\chi}_1^0},\, m_{\tilde{\tau}_1}$). 
Such UL maps allow us to directly set a limit on the corresponding topology appearing in a full model by interpolating the UL value to the given parameter point. However, their statistical interpretation is limited and only allows for a binary decision -- excluded or not -- on a topology-per-topology basis. 
Only if the expected ULs are provided in addition to the observed ones, it becomes possible to select the most sensitive result and/or to estimate an approximate likelihood for the signal strength. 

EMs are simulated acceptance times efficiency  ($A\times\epsilon$) values (simply called `efficiencies' in \smo) for the various signal regions of an analysis. Like the UL maps above, they are provided as grids in the simplified model parameter space. 
The big advantage of EMs is that they allow us to combine all contributions to a given signal region from different simplified model topologies. Furthermore, they enable the computation of a proper likelihood, either using a Gaussian approximation for the nuisance parameters
(simplified likelihood)~\cite{Ambrogi:2017neo,Ambrogi:2018ujg}, 
or using the full statistical model of the analysis~\cite{Alguero:2020grj} (see section~\ref{sec:pyhf}). 
They thus enable a richer and more robust statistical inference.

\begin{figure}[h!]
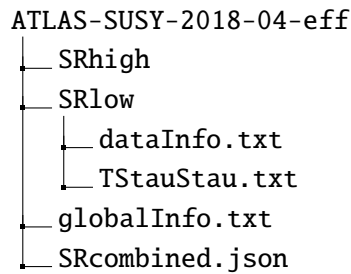

    \centering
    \begin{minipage}{4cm}
        \dirtree{%
        .1 ATLAS-SUSY-2018-04-eff. 
        .2 SRhigh. 
        .2 SRlow. 
        .3 dataInfo.txt. 
        .3 TStauStau.txt. 
        .2 globalInfo.txt. 
        .2 SRcombined.json. 
        }
    \end{minipage}
    \caption{File structure of the ATLAS-SUSY-2018-04~\cite{Aad:2019byo} EM  database entry}
    \label{fig:folder}
\end{figure}

An EM database entry is structured as shown in Figure~\ref{fig:folder}. 
It has sub-folders for each signal region, here {\tt SRhigh} and {\tt SRlow}, which contain the actual EMs, here {\tt TStauStau.txt}. Moreover, in each signal region sub-folder, there is a {\tt dataInfo.txt} file with relevant metadata, the number of observed and expected background events and the respective signal upper limits. (We will come back to the {\tt globalInfo.txt} and {\tt .json} files in section~\ref{sec:pyhf}.)

\noindent
UL database entries are structured analogously, although they do not have specific signal region sub-folders. 
The database text files are parsed and ``pickled'', using principal component analysis and Delaunay triangulation, to build a binary database file. 
The database folder and object structure as well as the binary (pickle) format are explained in detail in the \href{https://smodels.readthedocs.io/en/stable/DatabaseStructure.html}{database structure} section of the  \href{http://smodels.readthedocs.io/}{online manual}.
The whole database may be parsed and pickled locally, or downloaded as a ready-made pickle file.

\section{Latest database update}

In the v1.2.3 release~\cite{Khosa:2020zar}, the \smo\ database was updated with the results from \mbox{13 ATLAS} and \mbox{10 CMS} Run~2 analyses. In total, 76 official UL and EM results have been 
added in v1.2.3 (with 3 more EMs added in v1.2.4). This corresponds to all relevant new ATLAS and CMS results available on HEPData or on the analysis' TWiki page as of end of March 2020. An overview of these new results is given in Tables~\ref{tab:atlas} and \ref{tab:cms}. 

\begin{figure}[h]
    \centering
    \includegraphics[height=2cm]{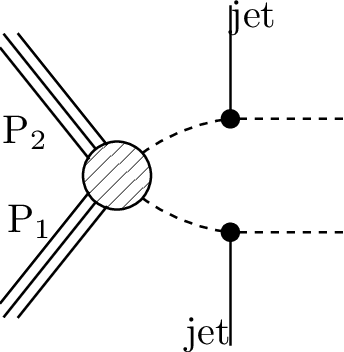}\qquad
    \includegraphics[height=2cm]{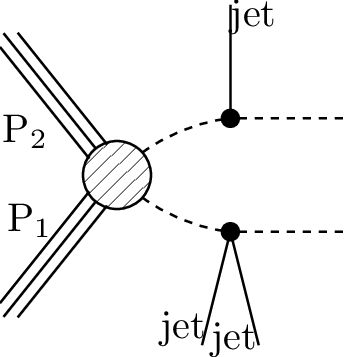}\qquad
    \includegraphics[height=2cm]{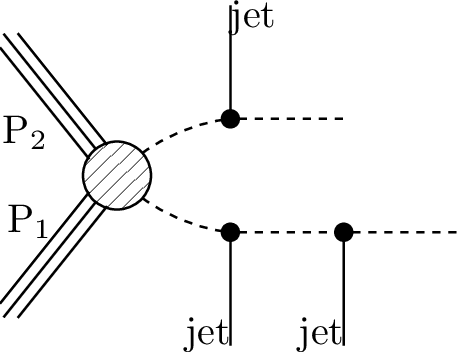}\qquad
    \includegraphics[height=2cm]{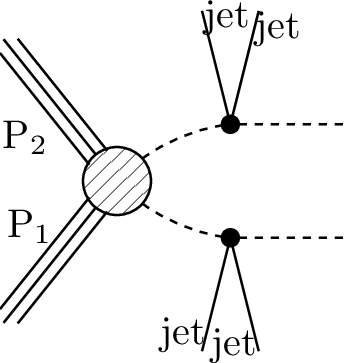}\qquad
    \includegraphics[height=2cm]{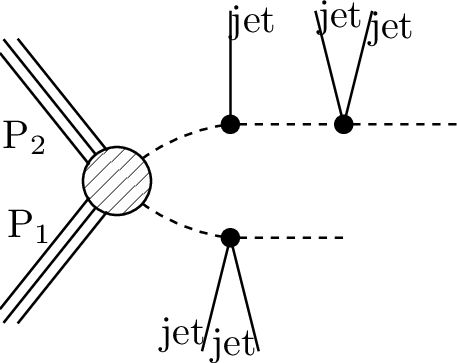}\\
    \vspace{5mm}
    \begin{tabular}{ lclcl }
        {\bf Analysis}  &  {\bf\bm $\cal{L}$ [fb$^{-1}$]}& \hphantom{SUSY}{\bf ID} & {\bf Ref.} & {\bf Type} \\ \hline
        0 lept.\ + jets  & 36.1 & ATLAS-SUSY-2016-07 &  \cite{Aaboud:2017vwy} & EM \\
        0 lept. + jets  & 35.9 &  CMS-SUS-16-033 & \cite{Sirunyan:2017cwe} & EM \\
        \hline
    \end{tabular}
    \caption{Topologies (top row) and analyses (bottom row) for which home-grown EM results were included in the recent database update.}
    \label{fig:hgrown}
\end{figure}

Moreover, 21 ``home-grown'' EMs were added for two ATLAS and CMS multi-jet + $\met$ searches at 13~TeV. These were produced by us using MadAnalysis\,5~\cite{Dumont:2014tja,Conte:2018vmg} to improve the coverage of topologies with 2--5 jets shown in Figure~\ref{fig:hgrown}; these are important for example when gluino-squark associated production is relevant.%
\footnote{Note here that home-grown EMs were included in \smo\ before, notably for stable heavy charged particles and R-hadrons~\cite{Heisig:2018kfq,Ambrogi:2018ujg}.}

Some more comments are in order.  
Regarding the new ATLAS analyses included in \smo, half of them have EMs in addition to the UL results, which as mentioned allows for the computation of a likelihood. It is also interesting to observe that five ATLAS analyses are at full Run~2 luminosity and three have full likelihoods available. 
On the CMS side, more than half of the official results have expected ULs in addition to the observed ones. However, none of them provides EMs. One CMS results is for full Run~2 luminosity.

\begin{table}[t]
\centering
\begin{tabular}{ lclcl }
    {\bf Analysis}  &  {\bf\bm $\cal{L}$ [fb$^{-1}$]} & \hphantom{SUSY}{\bf ID} & {\bf Ref.}  & {\bf Type} \\ \hline
    0 lept.\ + jets   & 36.1 & SUSY-2016-07 & \cite{Aaboud:2017vwy} & UL \\
    0 lept.\ stop & 36.1 & SUSY-2016-15 & \cite{Aaboud:2017ayj} &  UL \\
    1 lept.\ stop & 36.1 &  SUSY-2016-16 & \cite{Aaboud:2017aeu} &  UL, EM \\
    2--3 lept.\  & 36.1 &  SUSY-2016-24 & \cite{Aaboud:2018jiw} &  UL, EM \\
    photon + jets & 36.1 &  SUSY-2016-27 &  \cite{Aaboud:2018doq} & UL, EM \\
    0--1 lept. + $b$-jets  & 36.1 &  SUSY-2016-28 & \cite{Aaboud:2017wqg} &  UL \\
     EW-ino, Higgs & 36.1 & SUSY-2017-01 & \cite{Aaboud:2018ngk} & UL \\
     Higgsino, Z/H & 36.1 &  SUSY-2017-02 & \cite{Aaboud:2018htj} &  UL \\
     2 OS taus & 139.0 & SUSY-2018-04 & \cite{Aad:2019byo} & UL, EM, JSON \\
     3 lept., EW-inos & 139.0 &  SUSY-2018-06 & \cite{Aad:2019vvi} & UL$^\star$ \\
    multi-$b$ & 139.0 &  SUSY-2018-31 & \cite{Aad:2019pfy} &  UL, EM, JSON \\
     2 OS lept. & 139.0 &  SUSY-2018-32 & \cite{Aad:2019vnb} & UL \\
     1 lept. + $H\to b\bar b$ &  139.0 &  SUSY-2019-08 &  \cite{Aad:2019vvf} & UL, EM, JSON \\
    \hline
\end{tabular}
\caption{Official ATLAS 13~TeV results included in the recent database update. All analyses assume large $\met$ in the final state. The superscript $^\star$ denotes UL results which contain expected limits in addition to the observed ones. JSON denotes the availability of full likelihoods, see section~\ref{sec:pyhf}.}
\label{tab:atlas}
\end{table}

\begin{table}[ht]
\centering
\begin{tabular}{ lclcl }
    {\bf Analysis}  &  {\bf\bm $\cal{L}$ [fb$^{-1}$]}  & \hphantom{SUS}{\bf ID} &  {\bf Ref.} & {\bf Type} \\ \hline
    0 lept., top tagging   & 2.3 &  SUS-16-009 & \cite{Khachatryan:2017rhw} & UL$^\star$ \\
    $\geq 2$ taus & 35.9 &  SUS-17-003 &  \cite{Sirunyan:2018vig} & UL \\
    EW-ino combination  & 35.9 &  SUS-17-004 &  \cite{Sirunyan:2018ubx} & UL \\
    1 lept. compressed stop  & 35.9 &  SUS-17-005 & \cite{Sirunyan:2018omt} & UL$^\star$ \\
	jets + boosted $H\to b\bar b$  & 35.9 &  SUS-17-006 & \cite{Sirunyan:2017bsh} & UL$^\star$ \\
    2 SFOS lept. & 35.9 &  SUS-17-009 & \cite{Sirunyan:2018nwe} & UL$^\star$ \\
    2 lept. stop & 35.9 &  SUS-17-010 & \cite{Sirunyan:2018lul} & UL$^\star$ \\
    photon + ($b$) jets  & 35.9 &  SUS-18-002 &  \cite{Sirunyan:2019hzr} & UL$^\star$ \\
    0 lept. + jets, MHT    & 137.0 &  SUS-19-006 &  \cite{Sirunyan:2019ctn} & UL$^\star$ \\
    \hline
\end{tabular}
\caption{Official CMS 13~TeV results included in the recent database update. All analyses assume large $\met$ in the final state. The superscript $^\star$ denotes UL results which
contain expected limits in addition to the observed ones.}
\label{tab:cms}
\end{table}

\begin{figure}[ht]
    \centering
    \includegraphics[width=0.5\textwidth]{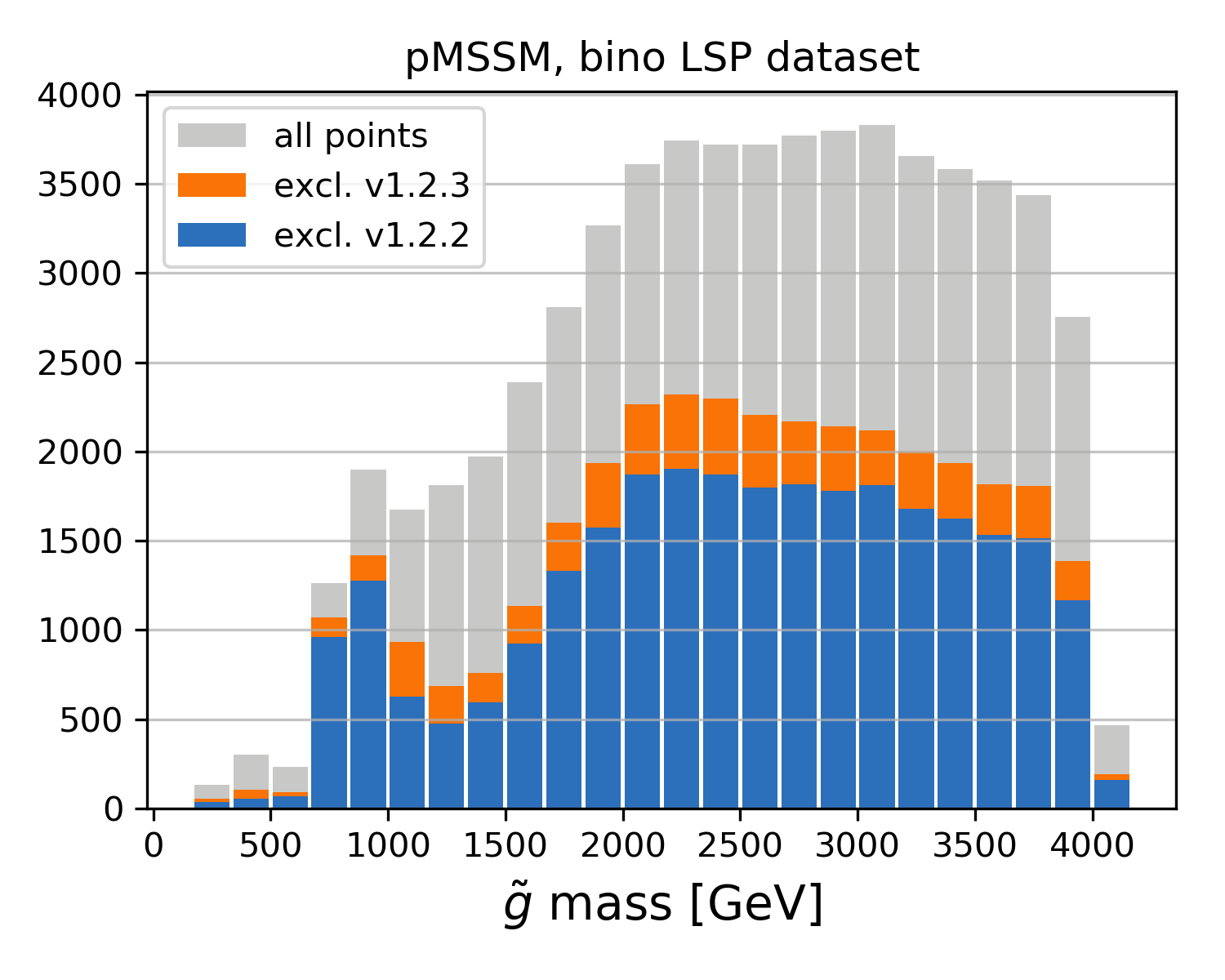}%
    \includegraphics[width=0.5\textwidth]{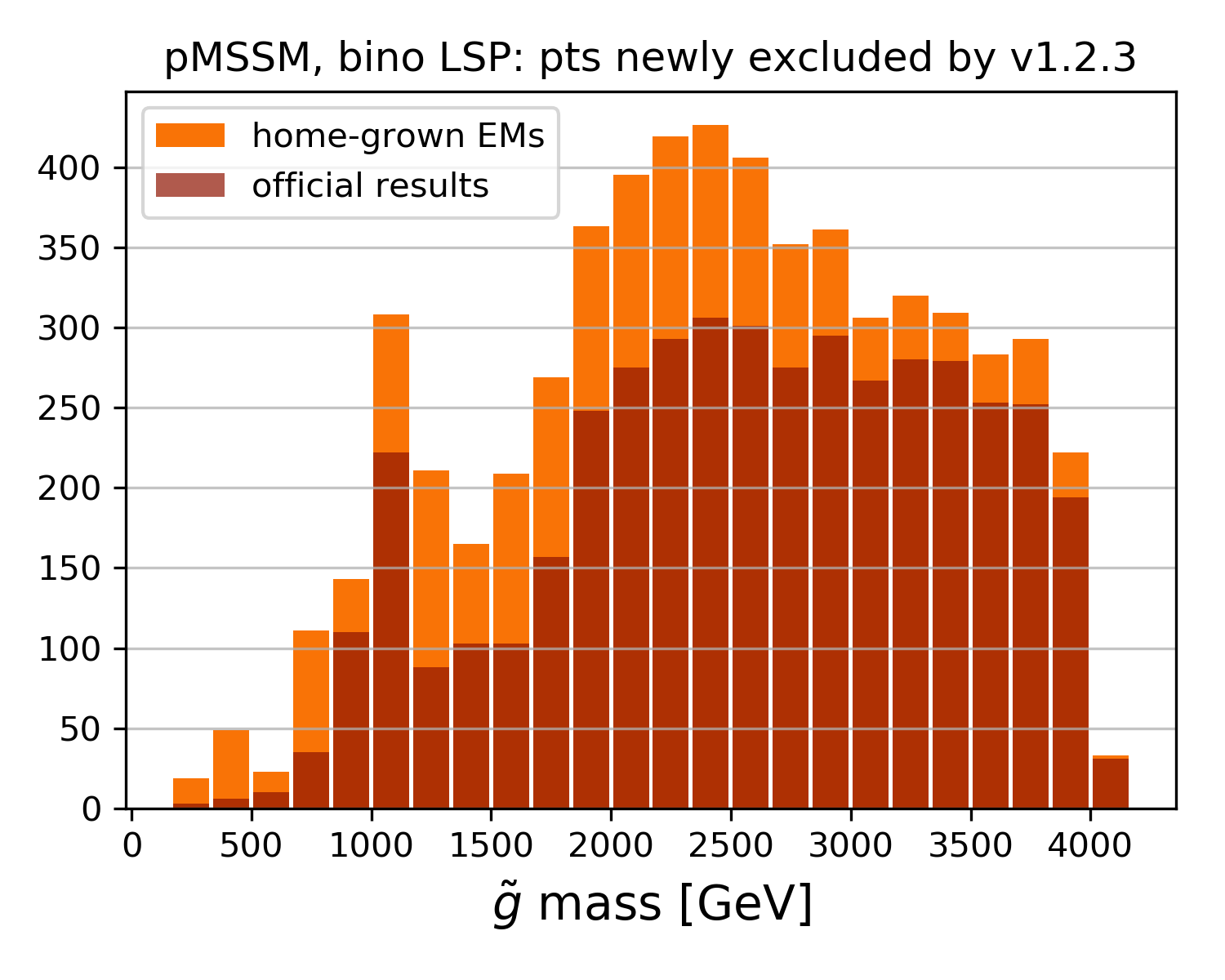}
    \caption{Results for the ATLAS pMSSM points with a bino-like LSP, which were classified as not excluded in \cite{Aad:2015baa}, as function of the mass of the gluino. 
    Left: In blue the number of points excluded in  version 1.2.2, in orange the number of points now excluded in v1.2.3 and in grey the total number of point per bin.
    Right: the number of newly excluded points when using only the official ATLAS and CMS results in v1.2.3 (dark red) as compared to using the full v1.2.3 database including the home-grown efficiency maps (orange).}
    \label{fig:dbUpdate}
\end{figure}

Figure~\ref{fig:dbUpdate} demonstrates the physics impact of 
the v1.2.3 database update upon the so-called phenomenological MSSM (pMSSM), i.e.\ the MSSM with 19 free parameters defined at the weak scale. This makes use of the extensive dataset from the ATLAS pMSSM study~\cite{Aad:2015baa} available at~\cite{ATLASpMSSMhepdata}; concretely we use the ``bino LSP'' points which are not excluded by the Run~1 searches considered in~\cite{Aad:2015baa}. 
The left plot in Figure~\ref{fig:dbUpdate} compares the marginal gluino mass distributions of points excluded with v1.2.3 to those excluded with v1.2.2. The plot on the right shows the number of newly excluded points when using only the official ATLAS and CMS results in v1.2.3 (dark red) as compared to using the full v1.2.3 database including the home-grown EMs (orange). 
The increase in constraining power due to the database update is quite noticeable. In numbers, \smo\ v1.2.3 excludes 10\% more points than v1.2.2, 27\% of which are due to the home-grown EMs. Additional plots showing the constraining power can be found in \cite{Khosa:2020zar}.

\section{Interface to \pyhf}\label{sec:pyhf}

As mentioned in section~\ref{sec:dbStruct}, EM-type results allow for the construction of a likelihood for the signal strength, and thus for a proper statistical evaluation, including, e.g., the computation of an  exclusion limit at a given CL\@. 
However, as long as no information on the correlation between signal regions is available, only the most sensitive (a.k.a.\ ``best'') signal region should be used for this exercise. This can severely limit the precision of the limit setting, see e.g.~contribution~15 of \cite{Brooijmans:2020yij}.
In turn, as demonstrated in e.g.~\cite{Asadi:2017qon},  being able to statistically combine disjoint signal regions instead of using only the best one, can be essential in physics studies.

The CMS SUSY group has been publishing signal region correlation data in the form of covariance matrices for some of their analyses.
This so-called simplified likelihood~\cite{CMS:2242860} approach assumes that uncertainties can be well approximated by Gaussians.   
\smo\ can make use of these correlation data since its version 1.2.0~\cite{Ambrogi:2018ujg}.

ATLAS has recently gone a significant step further by publishing \emph{full likelihoods} using a \json\ serialization~\cite{ATL-PHYS-PUB-2019-029}, which provides background estimates, changes under systematic variations, and observed data counts at the same fidelity as used in the experiment.
The \json\ format describes the \textsf{HistFactory} family of statistical models~\cite{Cranmer:1456844}, which is used by the majority of ATLAS searches.  The \pyhf\ package~\cite{pyhf} is then used to construct statistical models, and perform statistical inference, within a \python\ environment.%
\footnote{Note that this fulfills for the first time the Les Houches Recommendations 3b and 3c~\cite{Kraml:2012sg}!} 
The availability of such full likelihoods is a real boon for reinterpretation studies. From \smo v1.2.4 onwards, we are therefore providing an interface to \pyhf, which makes use of the ATLAS \json\ files whenever available~\cite{Alguero:2020grj}.  

As seen in Figure~\ref{fig:folder}, the \json\ file  describing the likelihood is placed in the folder of the associated database entry; in the example shown, the file is called {\tt SRcombined.json}. Note that this is a  background-only \json\ file, i.e.~it contains only background numbers with their uncertainties as well as experimentally observed numbers. Upon running \smo\ with signal region combination turned on, it is ``patched'' with the BSM contributions in the various signal regions as determined from the EMs. For this to work correctly, 
the name(s) of the \json\ file(s) to use, as well as the names of the relevant signal regions and their order, are specified in the {\tt globalInfo.txt} file.
(Apart from this, {\tt globalInfo.txt}  contains relevant metadata about the database entry, like the analysis ID, centre-of-mass energy, luminosity, etc.) 
In our example this is:
\begin{verbatim}
   datasetOrder: "SRlow", "SRhigh"
   jsonFiles: {"SRcombined.json": ["SRlow", "SRhigh"]}
\end{verbatim} 
\pyhf\ is then used for the statistical evaluation, returning a CL$_s$ that may be varied to infer a cross section upper limit at 95\% exclusion CL. 

\begin{figure}[t]
\centering
\includegraphics[width=0.5\textwidth]{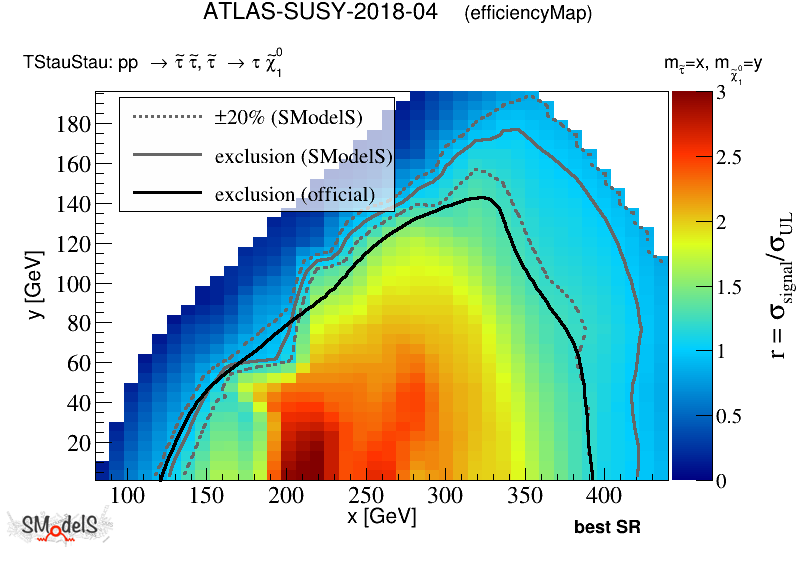}%
\includegraphics[width=0.5\textwidth]{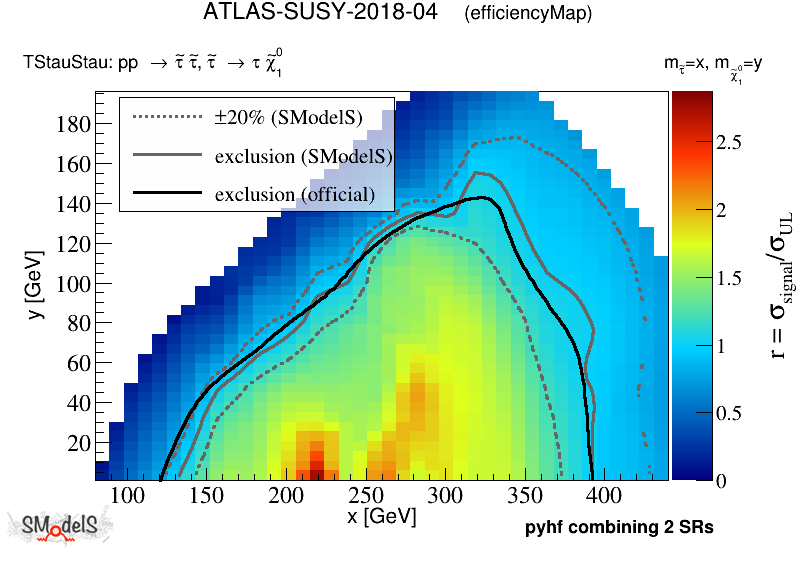}
\caption{Validation of the TStauStau ($pp\to \tilde\tau_1^+\tilde\tau_1^-$, $\tilde\tau^\pm_1\to\tau \tilde\chi^0_1$) result from the ATLAS stau search~\cite{Aad:2019byo}, on the left using the best signal region, on the right using the full likelihood.}
\label{fig:ATLAS-SUSY-2018-04}
\end{figure}

To demonstrate the physics impact, 
we compare in Figure~\ref{fig:ATLAS-SUSY-2018-04} the \smo\ exclusion (grey line) with the official exclusion (black line) for the ATLAS stau search~\cite{Aad:2019byo}, using the best signal region (left) and using \pyhf\ combination (right). As one can see, the usual procedure, which picks up the most sensitive efficiency map result, over-excludes by about 50 GeV on half the exclusion line. In contrast, a very good agreement with the official ATLAS result is obtained with the full \pyhf\ likelihood. The remaining small difference might be due to the (interpolated) acceptance $\times$ efficiency values from the simplified model EMs not exactly matching the ``true'' ones of the experimental analysis.

Even though we only show one result here, one can appreciate the gain in accuracy that one can reach with  full likelihoods. 
More examples can be found in \cite{Alguero:2020grj}.
The ATLAS collaboration is at the beginning of a huge effort to provide  full statistical models for new analyses. The first analyses published already show how this can help theorists make more trustful reinterpretations. The importance of such likelihood information for, e.g., global fits, has also been emphasised in~\cite{Abdallah:2020pec}.

\section{Installation and usage}

\noindent
\fbox{\parbox{0.98\linewidth}{%
	In this section we provide a short summary of the \href{https://indico.cern.ch/event/955391/contributions/4084223/attachments/2133802/3596360/smodels-gettingstarted.pdf}{quick guide for getting started} from the  \href{https://indico.cern.ch/event/955391/contributions/4084223/}{tutorial} given at this workshop. For more information, the user  is kindly referred to the  \href{http://smodels.readthedocs.io/}{online manual}, where a detailed documentation is available.}}\\[2mm]

\smo\ is a \python\ library that works with \python\ version 2.6 or later, the default being \python 3. 
It is listed in the Python Package Index (PyPI) and can thus be simply installed with {\tt pip} or {\tt pip3} 
by typing 
\begin{verbatim}
    pip install [--user] smodels
\end{verbatim}
This will install \smo\ into the default system or user directory, e.g.,~{\tt $\sim$/.local/lib/}. 
It is therefore recommended only for users who are at ease with working with \python\ libraries.

Most users may prefer to directly install the package in a local folder using the source file provided on the
\href{https://github.com/SModelS/smodels/releases}{\smo\ releases page} on github. To this end, download the  source ({\tt .tar.gz} or {\tt .zip}) file,
extract it at the desired destination and run: 
\begin{verbatim}
   make smodels
\end{verbatim}
in the top-level directory. This will install the required dependencies (using {\tt pip} install) and compile Pythia~\cite{Sjostrand:2014zea} and NLL-fast~\cite{Beenakker:2015rna}. For the latter two, a C++ and a Fortran compiler are needed. (If the Fortran compiler isn't found, try 
{\tt make FC=<path-to-gfortran> smodels}.)  
Details on install options and system-specific installations are given in the \href{https://smodels.readthedocs.io/en/latest/Installation.html}{installation section} of the  \href{http://smodels.readthedocs.io/}{online manual}. \\

The easiest way of using \smo\ is via the {\tt runSModelS.py} command-line tool. The basic format is
\begin{verbatim}
   ./runSModelS.py -f <path-to-input-file> [-p parameters.ini]   
\end{verbatim}
For the full list of options, type {\tt ./runSModelS.py -h}. 
The input file can be an SLHA or an LHE file. If a directory is given, all files in that directory will be processed. Note that input SLHA files need to include masses, decay tables, \emph{and LHC production cross sections}. (For adding cross sections for MSSM particles in SLHA format, \href{https://smodels.readthedocs.io/en/latest/SModelSTools.html}{\smotools} provides a convenient {\tt xsecomputer} based on Pythia and NLLfast.) 
A couple of sample input files for testing are provided in the {\tt inputFiles} directory. 

The basic options and parameters used by {\tt runSModelS.py} are defined in the \href{https://smodels.readthedocs.io/en/latest/RunningSModelS.html#parameterfile}{parameters file}. An example including all available parameters together with a short description, is stored in {\tt parameters.ini} in the top-level directory. 
Here also the signal region combination can be turned on or off by setting {\tt combineSRs} to True or False; as signal region combination may take a few seconds per point, False is the default. 

Two particularly important settings in the parameters file are the path to the database and the BSM model definition. For the database, one can provide either the path to the local text database, typically  
\begin{verbatim}
   path = ./smodels-database/
\end{verbatim}
in which case the database will be parsed and pickled locally, or an URL to the database pickle file. In the latter case, the binary pickle file will be downloaded from the server. This is often faster than pickling locally. 
Here, 
\begin{verbatim}
   path = official 
\end{verbatim}
provides a short-hand for the official database of your current \smo\ version. A list of public database versions is given on the \href{https://github.com/SModelS/smodels-database-release/releases}{database release page}.

The BSM model definition is done in a \python\ file (so-called particle module) that specifies the \emph{even} and \emph{odd} particle content of the model. In addition, the quantum numbers of the new particles must be given in a {\tt qNumbers} dictionary. A couple of pre-defined particle modules are available in {\tt smodels/share/models/}: {\tt mssm.py} and {\tt nmssm.py} for the MSSM and Next-to-MSSM, 
{\tt dgmssm.py} for the minimal Dirac gaugino model, and {\tt idm.py} for the Inert Doublet Model. In the parameters file, the path to the particle module is specified as, e.g., 
\begin{verbatim}
   model=share.models.idm 
\end{verbatim}
If no model is given, the MSSM is taken as the default.\\

While writing the particle module for a new model is rather straightforward, producing appropriate input SLHA files (including decay tables and cross sections) may be less so. Here the \textsf{micrOMEGAs}--\smo\ interface \cite{Barducci:2016pcb}, controlled with the switch 
\begin{verbatim}
   #define SMODELS 
\end{verbatim}
from \textsf{micrOMEGAs 4.3} onwards, 
provides a convenient way:  for all implemented models, including new ones added by the user, \textsf{micrOMEGAs} automatically produces the needed  {\tt particle.py} and SLHA-style \smo\ input files  for the parameter point under consideration (decay tables and cross sections are computed at tree level with \textsf{calcHEP}). \smo\ is called from  \textsf{micrOMEGAs} by the function
\begin{verbatim}
   smodels(Pcm, nf, csMinFb, fileName, wrt)
\end{verbatim}
where \verb|Pcm| is the proton beam energy in GeV and \verb|nf| is the number of parton flavors used to compute the production cross sections of the BSM  particles. The result is reported in SLHA format. 
The details of this functionality are described in \cite{Barducci:2016pcb} and in the manual shipped with \textsf{micrOMEGAs}.\\

This brings us to the \smo\ output. \smo\ primarily reports its results in the form of {\bf $\bm r$-values}, defined as the ratio of the theory prediction over the observed upper limit, for each experimental constraint that is matched in the database. Generally, all points for which at least one $r$-value equals or exceeds unity ($r_{\rm max} \ge 1$) are considered as excluded. 
For EM-type results, when {\tt computeStatistics=True}, the likelihood and $\chi^2$ are also reported.  
As an example, Figure~\ref{fig:outputExample} shows part of the output obtained for the {\tt gluino\_squarks.slha} sample input file located in {\tt inputFiles/slha/}. 

\begin{figure}[]{\small
\begin{verbatim}
Input status: 1
Decomposition output status: 1 #decomposition was successful
# Input File: inputFiles/slha/gluino_squarks.slha
# maxcond = 0.2
# minmassgap = 5.
# ncpus = 1
# sigmacut = 0.01
# Database version: 1.2.4
================================================================================
#Analysis  Sqrts  Cond_Violation  Theory_Value(fb)  Exp_limit(fb)  r  r_expected

 ATLAS-SUSY-2016-07  1.30E+01    0.0  9.791E+00  1.270E+00  7.710E+00  9.151E+00
 Signal Region:  5j_Meff_1600
 Txnames:  T1, T2, T5GQ, T5WW, T5ZZ, T6WW, TGQ
 Chi2, Likelihood =  2.577E+01  2.469E-09
--------------------------------------------------------------------------------
     CMS-SUS-19-006  1.30E+01    0.0  1.169E+01  5.138E+00  2.274E+00  1.898E+00
 Signal Region:  (UL)
 Txnames:  T2
--------------------------------------------------------------------------------
....
....
================================================================================
The highest r value is = 7.709501308305477
\end{verbatim} } 
\caption{Part of the summary output ({\tt gluino\_squarks.slha.smodels} file) obtained when processing 
{\tt ./runSModelS.py -f inputFiles/slha/gluino\_squarks.slha -p parameters.ini}. This is an MSSM scenario with $m_{\tilde g}=865$~GeV and $m_{\tilde q}\approx 990$~GeV. The results constraining this scenario are reported from highest to lowest  $r$-value. In this example, the highest $r$-value comes from the home-grown EMs for the ATLAS-SUSY-2016-07 analysis (see Figure~\ref{fig:hgrown}), combining the contributions from several signal topologies. The second-highest $r$-value comes from the UL result for the T2 topology from CMS-SUS-19-009 (see Table~\ref{tab:cms}). The ``Txnames'' nomenclature for the topologies is explained in the \href{https://smodels.github.io/docs/SmsDictionary}{SMS dictionary}. 
Note that for the EM result, the $\chi^2$ and likelihood are reported in addition to the observed and expected $r$-values.  
The full output also contains information on unconstrained cross sections (missing topologies and outside grid elements), not displayed here.}
    \label{fig:outputExample}
\end{figure}

Moreover, if {\tt testCoverage=True}, \smo\ prints a list of {\bf missing topologies}:  elements which are not tested by any of the experimental results in the database (independent of the element mass). 
Finally, {\bf outside grid} elements, which could be tested by one or more experimental result, but are not constrained because the mass array is outside the mass grid, are reported.

Different output formats are available: summary, stdout, log, python, xml, slha. They are chosen and controlled from the {\tt parameters.ini} file. 
A detailed explanation of the information contained in each type of output is given in the \href{https://smodels.readthedocs.io/en/latest/OutputDescription.html#outputdescription}{SModels Output} section of the  \href{http://smodels.readthedocs.io/}{online manual}. \\

Although {\tt runSModelS.py} provides the main \smo\ features with a command line interface, users more familiar with \python\ and the \smo\ language may prefer to write their own main program. A simple, commented example code for this purpose is provided as {\tt Example.py} in the \smo\ distribution. 
The \href{https://smodels.readthedocs.io/en/latest/Examples.html}{How To's} in the  \href{http://smodels.readthedocs.io/}{online manual} provide additional examples for using \smo\ and some of the \smotools\ as a \python\ library.\\ 

\noindent
Last but not least, users can contact the \smo\ developers at  
\begin{verbatim}
   smodels-users@lists.oeaw.ac.at 
\end{verbatim}
for any questions or comments which may arise using the package. New contributors are also always welcome.

\section{Summary and outlook to SModelS 2.0}
\label{summary}

We gave a short introduction to constraining new physics with \smo\ 
and presented two important updates of the tool published in 2020: \emph{(i)}~a major database update with results from 13 ATLAS and 10 CMS SUSY searches at 13~TeV, and \emph{(ii)}~the interface to \pyhf, which enables \smo\ to use full likelihoods from ATLAS. The database update significantly increases the constraining power of \smo\ and the ability to use full likelihood models leads to a marked gain in accuracy (whenever full likelihoods are available).  

While the examples discussed in this contribution focus on $\met$ final states, 
the capabilities of \smo\ go beyond $\met$ searches for new physics. 
Indeed, long-lived particles (LLPs) constrained by searches for heavy stable charged particles (HSCP) or $R$-hadrons, can be treated in \smo\ since version 1.2~\cite{Heisig:2018kfq,Ambrogi:2018ujg}. These searches require the LLP to decay outside the (relevant parts of the) detector, which introduces a dependence on the particle's total decay width. To apply such results to intermediate lifetimes, which lead to a fraction of the decays happening inside the detector, \smo\ computes the fractions of prompt decays and decays taking place outside the detector 
assuming a constant boost factor for the LLP~\cite{Heisig:2018kfq}.
The impact of the currently implemented HSCP searches has recently been demonstrated for the Dirac gaugino case in \cite{Goodsell:2020lpx}, where it was shown that their constraining power is at the level of the full recasting approach. 

Ongoing developments include 
a major revision of the decomposition algorithm that enables the support of general LLP signatures. Most significantly, it introduces the particles' decay widths and their quantum numbers as additional parameters of the simplified model topologies. Accordingly, the experimental results in the database can be width dependent. This will enable to incorporate a wide range of exotic signatures, such as searches for disappearing and kinked tracks, displaced jets and leptons as well as delayed jets and photons. This is especially relevant for LLPs with decay lengths of the order of millimeters to a few meters.

The inclusion of the particle quantum numbers in the SMS topologies will provide further opportunities. 
For instance, it will permit to specify the applicability of certain experimental results to BSM particles with specific spins, allowing to go beyond the assumptions mentioned in footnote~\ref{fn:assump}. 
Finally, the particle content of full models will be definable through the use of an input SLHA file with QNUMBERS blocks, which will greatly simplify using \smo\ for models other than the MSSM\@.
These features will be available with the upcoming \smo\ version 2.0.

\section*{Acknowledgements}
The work of G.A.\ and S.K.\ is supported in part by the IN2P3 project ``Th\'eorie -- BSMGA''. J.H.~acknowledges support from the F.R.S.-FNRS, of which he is a postdoctoral researcher. Su.K. ~acknowledges support from the FWF Elise-Richter fellowship under project number V592-N27.


\providecommand{\href}[2]{#2}\begingroup\raggedright\endgroup

\end{document}